\documentclass[10pt,a4paper,aps,pra,twocolumn,superscriptaddress,showpacs]{revtex4-1}

\usepackage{amsbsy}
\usepackage{amsfonts}
\usepackage{amsmath}
\usepackage{amssymb}
\usepackage{array}
\usepackage{bbding}
\usepackage{bbm}
\usepackage{bm}
\usepackage{epsfig}
\usepackage{fixmath}
\usepackage[T1]{fontenc}
\usepackage{graphicx}
\usepackage{multirow}
\usepackage{ogonek}
\usepackage[svgnames]{xcolor}
\usepackage{tikz}

\providecommand*{\I}{\mathrm{i}}                           
\renewcommand{\vec}[1]{\mathbold{#1}}
\newcommand{\spinor}[3]{\left(\hspace{-1.0ex}\begin{array}{c}#1\\ #2\end{array}\hspace{-1.0ex}\right)_{\hspace{-0.7ex}#3}\hspace{-2.0ex}}

\newcommand{\nn}{\nonumber}
\renewcommand{\d}{\mathrm{d}}
\newcommand{\ie}{i.\hspace{0.08em}{\nolinebreak}e.$\left.\right.$\hspace{-0.3em}}
\newcommand{\eg}{e.\hspace{0.08em}{\nolinebreak}g.$\left.\right.$\hspace{-0.3em}}

\newcommand{\wrt}{w.\hspace{0.08em}{\nolinebreak}r.\hspace{0.08em}{\nolinebreak}t.$\left.\right.$}

\newcommand{\mw}[1]{\left<\right.\hspace{-0.5ex}{#1}\left.\hspace{-0.5ex}\right>}

\newcommand{\nab}{\mathbold{\nabla}}

\definecolor{ao(english)}{rgb}{0.0, 0.5, 0.0}

\begin{document}


\title{Ground-state densities of repulsive two-component Fermi gases}


\author{Martin-Isbj\"orn Trappe}
\email[]{martin.trappe@quantumlah.org}
\affiliation{Center for Theoretical Physics PAN, Al.~Lotnik\'ow 32/46, 02-668 Warsaw, Poland}
\affiliation{Centre for Quantum Technologies, National University of Singapore, Block S15, 3 Science Drive 2, Singapore 117543}

\author{Piotr Grochowski}
\email[]{p.t.grochowski@gmail.com}
\affiliation{Center for Theoretical Physics PAN, Al.~Lotnik\'ow 32/46, 02-668 Warsaw, Poland}

\author{Miros{\l}aw Brewczyk}
\email[]{m.brewczyk@uwb.edu.pl}
\affiliation{Center for Theoretical Physics PAN, Al.~Lotnik\'ow 32/46, 02-668 Warsaw, Poland}
\affiliation{Wydzia{\l} Fizyki, Uniwersytet w Bia{\l}ymstoku, ul. K. Cio{\l}kowskiego 1L, 15-245  Bia{\l}ystok, Poland}

\author{Kazimierz Rz\k{a}$\dot{\mbox{z}}$ewski}
\email[]{kazik@cft.edu.pl}
\affiliation{Center for Theoretical Physics PAN, Al.~Lotnik\'ow 32/46, 02-668 Warsaw, Poland}



\date{\today}

\begin{abstract}
We investigate separations of trapped balanced two-component atomic Fermi gases with repulsive contact interaction. Candidates for ground-state densities are obtained from the imaginary-time evolution of a nonlinear pseudo-Schr\"odinger equation in three dimensions, rather than from the cumbersome variational equations. With the underlying hydrodynamical approach, gradient corrections to the Thomas-Fermi approximation are conveniently included and are shown to be vital for reliable density profiles. We provide critical repulsion strengths that mark the onset of phase transitions in a harmonic trap. We present transitions from identical density profiles of the two fermion species towards isotropic and anisotropic separations for various confinements, including harmonic and double-well-type traps. Our proposed method is suited for arbitrary trap geometries and can be straightforwardly extended to study dynamics in the light of ongoing experiments on degenerate Fermi gases. 
\end{abstract}


\maketitle


\section{\label{Intro}Introduction}

In recent years ultracold Fermi gases have come into focus to study and manipulate novel phases of matter that are less accessible in strongly interacting many-body systems such as solid states or stellar matter. For example, two-component Fermi gases with repulsive short-range interactions may exhibit magnetic properties like transitions from para- to ferromagnetic phases. Considerable efforts in both experiment and theory have been made to determine static and dynamic properties of such systems \cite{DeMarco2002,Du2008,Ketterle2009,Roati2011,Ketterle2012}. For instance, itinerant ferromagnetism in strongly repulsive Fermi gases has been extensively investigated \cite{Kim2004,Kim2005,Duine2005,Ketterle2009,Conduit2009,Roati2011,Ketterle2012,Troyer2013arxiv}. But both experimental and theoretical results on separations of the two fermion species are ambiguous and a number of fundamental questions, in particular away from unitarity, remain elusive to date.

Dynamics of two strongly interacting clouds of different components and dynamical tunneling through barriers like in \cite{Roati2011} are, for example, addressed with a hydrodynamical approach in Thomas-Fermi (TF) approximation \cite{DissSchneider,Schneider2011}. The TF-approximation, first developed for multi-electron atoms \cite{Thomas1927,Fermi1927}, is frequently used to obtain a first approximate description of a many-fermion system. However, this semiclassical local density approximation may fail if an accurate description in classically forbidden regions is required. Then, corrections in terms of gradients of the particle density are usually taken into account \cite{Kirzhnits1957,Hodges1973,Murphy1981,Holas1991,Salasnich2007,Koivisto2007,Salasnich2008,Csordas2010}. It can be expected that such gradient corrections also play a crucial role at interfaces of two repelling fermionic clouds.

The regime of weak repulsion is even less explored. In specific settings, gradient corrections are reported to be negligible for the total density, see for example \cite{Salasnich2013}. But little is known about the ground-state densities of the individual fermion components for large particle numbers beyond the TF-approximation, even for the simple case of a repulsive contact interaction, which is addressed in the present work. Of course, other interactions may have to be included, depending on the specific physical setting. For example, dipole-dipole-interacting Fermi gases are addressed in \cite{Kazik2003,FangBerge2011,Kazik2013}.

While few-fermion systems permit exact diagonalization, restricted computing resources require approximate methods for large particle numbers, in particular in three spatial dimensions. Separations and domain structures for up to $100$ particles in one spatial dimension were found in \cite{Kazik2004} with the aid of a mean-field Hartree-Fock method. The phase-diagram for short-range hard/soft-sphere interaction in a flat box and for small particle numbers is addressed in \cite{Troyer2010}. For large particle numbers the TF-approximation with contact interaction, employed in \cite{Salasnich2000,Tosi2000,Sogo2002,XuGu2011,Sun2014}, yields (partial) separations of the two fermion components. LeBlanc~et.~al.~compared spin textures, included gradient corrections, and reported an isotropic separation in case of a fixed quantization axis in a harmonic trap \cite{LeBlanc2009}.

In this work we provide an extensive characterization of ground-state density candidates for two-component Fermi gases with repulsive contact interaction in three dimensions for various trapping potentials. We want to answer the question to what extent the TF-approximation and its extensions predict separations of the two Fermi components. In Sec.~\ref{Energydensityfunctional} we discuss the approximate energy functional, including gradient corrections, that underlies the variational equations which are commonly employed to obtain ground-state densities in the spirit of density functional theory. We argue in Sec.~\ref{InadequacyTF} that the variational equations without gradient corrections, viz., the TF equations predict ground states candidates ambiguously. In Sec.~\ref{Quantumhydro} we establish a pseudo-Schr\"odinger equation for three spatial dimensions, whose propagation in imaginary time yields stationary states. This approach is based on Madelung's hydrodynamical equations and allows us to conveniently study the qualitative and quantitative effect of the gradient corrections in arbitrary trap geometries. In contrast, the variational equations with gradient corrections included are in general anisotropic coupled nonlinear partial differential equations in three dimension, and their direct solution is cumbersome. Our results for particle densities in various potentials obtained from the imaginary-time evolution method are summarized in Sec.~\ref{ImagTimeEvol}, where we discuss the observed types of separation. For harmonic confinement we map out the phase diagram of critical interaction strengths at which phase transitions towards separation occur and present numerous density profiles, also for trapping potentials with tunneling barriers. We explicitly address the qualitative and quantitative importance of gradient corrections beyond the TF approximation, which is briefly discussed in the appendix. Throughout this work we use harmonic oscillator units and set ${\hbar=\omega=m=1}$.

\section{Energy density-functional}\label{Energydensityfunctional}

We base our investigation on the total energy ${E=E_{\mathrm{kin}}+E_{\mathrm{pot}}+E_{\mathrm{int}}}$, composed of kinetic, potential, and interaction energy. In Thomas-Fermi (TF) approximation the kinetic energy is replaced by the Thomas-Fermi kinetic energy density-functional for $D$ spatial dimensions,
\begin{align}\label{TTF}
T_{\mathrm{TF}}[n_+,n_-]=c_D\int(\d\vec r)\left\{(n_+)^\frac{D+2}{D}+(n_-)^\frac{D+2}{D}\right\},
\end{align}
depending on the one-particle densities $n_+(\vec r)$ and $n_-(\vec r)$ of the two fermion species. For brevity we do not indicate the $\vec r$-dependence of quantities where appropriate. The coefficients $c_D$ for spin-polarized fermions read ${c_1=\hbar^2\pi^2/(6m)}$, ${c_2=\hbar^2\pi/m}$, and ${c_3=6^{5/3}\hbar^2\pi^{4/3}/(20m)}$. Commonly, gradient corrections in terms of a formal ${\hbar\mbox{-expansion}}$ are added to the TF kinetic energy functional to improve on the TF approximation. Given slowly varying potentials, the first-order corrections in three dimensions (3D) are
\begin{align}\label{DeltaT}
\Delta T_{\hbar^2}[n_+,n_-]&=\int(\d\vec r)\frac{\xi\hbar^2}{8m}\left\{\frac{(\nab n_+)^2}{n_+}+\frac{(\nab n_-)^2}{n_-}\right\},
\end{align}
\ie, $E_{\mathrm{kin}}$ is approximated by ${T_\xi=T_{\mathrm{TF}}+\Delta T_{\hbar^2}}$. With respect to a systematic treatment we adhere to
\begin{align}\label{xi}
\xi=\frac19,
\end{align}
see for example \cite{Kirzhnits1957}. Of course, if higher order gradient corrections are omitted, other values of $\xi$ may yield a better approximation of the true kinetic energy, depending on the physical situation. 

The orbital-free expressions of the gradient corrections in 1D and 2D presented in the literature are troublesome: The 1D gradient corrections are not bounded from below, and vanishing corrections (to all orders of $\hbar$) are reported for 2D \cite{Holas1991,Salasnich2007,Koivisto2007,Putaja2012}, that is, the employed methods are inappropriate for low-dimensional systems. In contrast, the gradient corrections in 3D are consistently derived with various methods and exhibit no obvious shortcomings \cite{Kirzhnits1957,Hodges1973,Murphy1981,Holas1991,Salasnich2007,Koivisto2007}. We shall therefore limit our discussion of gradient corrections to the 3D case.

The potential energy of the two fermion species in their respective external potentials $V_\pm(\vec r)$ is
\begin{align}\label{Epot}
E_{\mathrm{pot}}=\int(\d\vec r)\,(V_+\,n_++V_-\,n_-),
\end{align}
and the repulsive contact interaction energy is
\begin{align}\label{Eint}
E_{\mathrm{int}}=g\int(\d\vec r)\,n_+\,n_-
\end{align}
with interaction strength ${g\ge0}$, related to the $s$-wave scattering length $a_s$ through ${g=4\pi\,a_s\hbar^2/m}$. Hence, we choose the total energy functional
\begin{align}\label{Etot}
E_\xi[n_+,n_-,\mu_+,\mu_-]&=T_\xi+E_{\mathrm{pot}}+E_{\mathrm{int}}\nn\\
&\quad+\sum_{j=\pm}\mu_j\left(N_j-\int(\d\vec r)\,n_j\right)
\end{align}
for unrestricted minimization over $n_\pm(\vec r)$ and $\mu_\pm$ in the spirit of density functional theory. The conservation of the particle numbers $N_+$ and $N_-$ is enforced through Lagrange multipliers $\mu_+$ and $\mu_-$, \ie, the chemical potentials of the two fermion species. The choice $\xi=0$ in (\ref{Etot}) corresponds to the TF approximation of $E_{\mathrm{kin}}$. In the limit ${g\to0}$ the energy (\ref{Etot}) reduces to that of two independent Fermi gases.

The contact interaction term (\ref{Eint}) is obtained from second quantization, where the part of the Hamiltonian responsible for two-body interactions is written in the form
\begin{align}
V_{\mathrm{int}} = \frac{1}{2}  \sum_{i,j=\pm} \int(\d\vec r)(\d\vec r')\Big\{& \hat{\psi}^{\dagger}_i(\mathbf{r})\hat{\psi}^{\dagger}_j(\mathbf{r}') V^{ij}_{\mathrm{int}}(\mathbf{r},\mathbf{r}')\nn\\
&\times\hat{\psi}_j(\mathbf{r}') \hat{\psi}_i(\mathbf{r})\Big\}
\label{2body}
\end{align}
with a two-body interaction potential
\begin{align}
V^{ij}_{\mathrm{int}}(\mathbf{r},\mathbf{r}') = g\, \delta(\mathbf{r}-\mathbf{r}')\, (1 - \delta_{ij}). 
\label{pot}
\end{align}
The summation in (\ref{2body}) yields
\begin{align}
V_{\mathrm{int}} &= \frac{1}{2}  \,g
\int(\d\vec r)\, \hat{\psi}^{\dagger}_{+}(\mathbf{r})
\hat{\psi}^{\dagger}_{-}(\mathbf{r}) 
\hat{\psi}_{-}(\mathbf{r}) \hat{\psi}_{+}(\mathbf{r})  \nn  \\
&\quad + \frac{1}{2}  \,g
\int(\d\vec r)\, \hat{\psi}^{\dagger}_{-}(\mathbf{r})
\hat{\psi}^{\dagger}_{+}(\mathbf{r}) 
\hat{\psi}_{+}(\mathbf{r}) \hat{\psi}_{-}(\mathbf{r}).
\label{2body1}
\end{align}
Introducing the density operators $\hat{n}_{\pm}(\mathbf{r})= \hat{\psi}^{\dagger}_{\pm}(\mathbf{r}) \hat{\psi}_{\pm}(\mathbf{r})$ and assuming that both components are highly occupied (which allows us to replace the density operators with real functions), one eventually obtains
\begin{align}
V_{\mathrm{int}} = g\int(\d\vec r)\,  n_{+} n_{-},
\label{final}
\end{align}
as already displayed in (\ref{Eint}). What is left out in (\ref{final}) can be seen by calculating the average of the product of the four field operators that appear under the integrals in (\ref{2body1}). For an ideal gas, based on the Wick's theorem, we have
\begin{align}
\mw{\hat{\psi}^{\dagger}_{+} \hat{\psi}^{\dagger}_{-} \hat{\psi}_{-} \hat{\psi}_{+}} &=
\mw{\hat{\psi}^{\dagger}_{+} \hat{\psi}_{+}}\mw{\hat{\psi}^{\dagger}_{-} \hat{\psi}_{-}} - \mw{\hat{\psi}^{\dagger}_{+} \hat{\psi}_{-}} \mw{\hat{\psi}^{\dagger}_{-} \hat{\psi}_{+}} \nn  \\
&\quad + \mw{\hat{\psi}^{\dagger}_{+} \hat{\psi}^{\dagger}_{-}} \mw{\hat{\psi}_{-} \hat{\psi}_{+}}
\label{Wick}
\end{align}
for both integrands. The third term on the right-hand side of (\ref{Wick}) vanishes since the total number of atoms is preserved. Now, it is clear that (\ref{final}) differs from (\ref{2body1}) by terms like the second one in the right-hand side of (\ref{Wick}) --- terms which describe the intercomponent correlations. Obviously, they vanish in the limit of an ideal gas. We neglect these terms throughout this work and employ $g$ as a free parameter.

The global minimum of the approximate energy functional (\ref{Etot}) is attained by approximate ground state densities. A standard approach for finding $n_\pm$ is to solve the variational equations obtained from the stationarity condition ${\delta E_\xi=0}$. In 3D, functional differentiation of (\ref{Etot}) \wrt the particle densities $n_\pm$ leads to
\begin{align}\label{TFvWeq}
\frac53 c_3\,n_\pm^{2/3}-\xi\frac{\hbar^2}{2m}\frac{\nab^2\sqrt{n_\pm}}{\sqrt{n_\pm}}+V_\pm-\mu_\pm+g\,n_\mp=0.
\end{align}

Commonly, the gradient corrections are omitted and the resulting TF energy-functional is minimized to get a first approximation of the fermionic clouds. To estimate the quality of the TF approximation, one has to go beyond the TF equations. But including the gradient corrections in (\ref{TFvWeq}), we face coupled partial differential equations in 3D that are tedious to solve in anisotropic situations. Differentiable densities are included at the level of the energy functional via gradient corrections which turn out to be necessary to obtain viable ground state densities even qualitatively and for large particle numbers.

\section{Thomas-Fermi equations}\label{InadequacyTF}

Omitting gradient corrections (${\xi=0}$) and employing the same external potential ${V_\pm=V}$ for both fermion species, we obtain the algebraic TF equations
\begin{align}\label{TFeq}
A\,n_\pm^{2/D}+g\,n_\mp=\mu_\pm-V
\end{align}
from the variation of (\ref{Etot}), where ${A=c_D(D+2)/D}$. The TF equations were addressed and used in several publications \cite{Tosi2000,Salasnich2000,Sogo2002,LeBlanc2009,XuGu2011,Sun2014}. 

The TF density profiles are ambiguous in two ways. They do not need to be continuous, since solutions of (\ref{TFeq}) at different positions $\vec r$ are decoupled. Consequently, the decoupling of directions makes the identification of anisotropic separations of $n_+$ and $n_-$ ambiguous. There can be several pairs $\{n_+(\vec r),n_-(\vec r)\}$ of solutions for a given $\vec r$. This potentially leaves us with a myriad of density profiles that can be very close in energy since the various density profiles can be chopped up and recombined in arbitrarily many ways. Moreover, since the TF densities generally can yield energies that differ from the exact energies by a few percent, one should not be too confident in preferring one density profile over another if both yield only slightly different energies. 

\begin{figure}[h!tbp]
\centering
\includegraphics[width=0.99\linewidth]{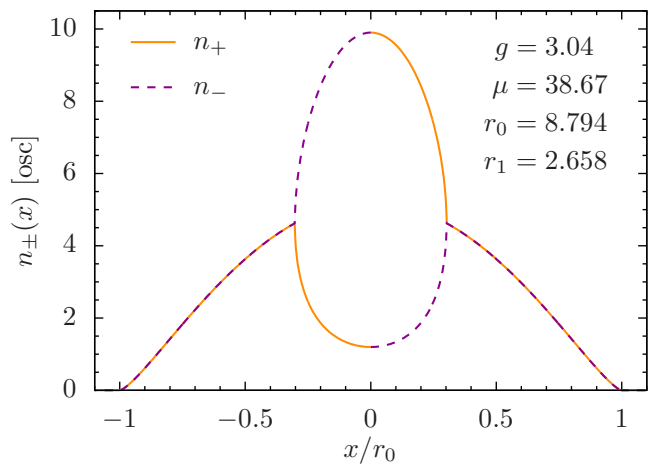}
\caption{(Color online) Illustration of the generic structure of radial densities $n_\pm(r)$, depicted along the $x$-axis, for ${\mu_\pm=\mu}$ and ${N_\pm=5000}$ particles in 3D. All quantities are given in harmonic oscillator units [osc]. The displayed separated density profile has marginally lower energy than the symmetric solution: ${E_{0,\mathrm{sep.}}=287915<E_{0,\mathrm{symm}}=288487}$. The radius $r_0$ at which both densities $n_\pm$ vanish is given by ${V(r_0)=\mu}$. The radius $r_1$ where the separated solutions merge with the symmetric solutions is given in the appendix.}
\label{TwoFermiComponentsPlotsn3D}
\end{figure}

\begin{figure}[h!tbp]
\centering
\includegraphics[width=0.99\linewidth]{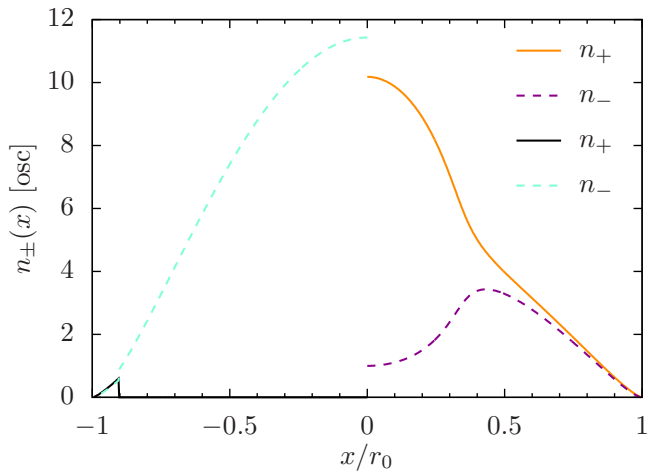}
\caption{(Color online) Possible density profiles along the $x$-axis for ${\mu_+=38.722}$, ${\mu_-=38.545}$, and ${g=3.04}$, selected from two (of many discontinuous) radial TF-solutions. Here, we choose ${r_0=\sqrt{2\mu_+/(m\omega^2)}}$.}
\label{TwoFermiComponentsPlotsn3Dmupm}
\end{figure}

One option to deal with these ambiguities is to restrict the TF solutions to those that are continuous, although, strictly speaking, this condition is not justified for the (spatially decoupled) solutions of (\ref{TFeq}). Furthermore, for a first estimate of the particle densities it can be sufficient to consider ${\mu_+=\mu_-}$ like in \cite{XuGu2011}. Then, there are analytical solutions of (\ref{TFeq}) in all dimensions. In particular, we find symmetric solutions ${n_+(\vec r)=n_-(\vec r)}$ for all $\vec r$. Details are discussed in the appendix. However, those solutions may be regarded as reasonable ground-state candidates only for selected parameters. For example, in Fig.~\ref{TwoFermiComponentsPlotsn3D} we depict the generic radial density profiles obtained from (\ref{TFeq}) in the 3D isotropic potential ${V(\vec r)=\frac12 \vec r^2}$ for a repulsion strength $g$ that allows for (partial) separation of the two Fermi components. The identical chemical potentials allow us to swap the roles of $n_+(\vec r)$ and $n_-(\vec r)$ at any $\vec r$ if we do not impose continuity of the densities. We may then disregard continuity, use the radial densities for ${x>0}$, and swap the roles of $n_\pm$ for ${x<0}$ to end up with ${N_\pm=5000}$. Continuity would require ${N_+\not=N_-}$.

In general, the TF densities are obtained numerically for any particle numbers and for ${\mu_+\not=\mu_-}$, see \cite{Salasnich2000} for an early account. Figure~\ref{TwoFermiComponentsPlotsn3Dmupm} illustrates examples of radial densities for ${N_\pm=5000}$ and ${\mu_+\not=\mu_-}$. The densities can be chopped up and recombined such that we arrive at ${N_\pm=5000}$ on various routes. Among all the so obtained density distributions the one that yields minimal energy $E_{0,\mathrm{min}}$ would be our ground state candidate in the spirit of density functional theory. It is clear, however, that there are many (not necessarily similar density distributions) with energies very close to $E_{0,\mathrm{min}}$. A rigorous selection of the ground state densities is therefore a questionable undertaking. The TF equations may suffice for a decent estimate of the energy, although large density gradients like those at the interfaces of the different radial solutions in Fig.~\ref{TwoFermiComponentsPlotsn3D} and \ref{TwoFermiComponentsPlotsn3Dmupm} can be expected to require corrections to the kinetic energy. 

In summary, realistic densities are not readily obtainable from (\ref{TFeq}) since we cannot get rid of the aforementioned ambiguities within the realm of the TF equations. An approach that enforces continuity is called for, and a natural way is to include density gradients. In turn, density profiles obtained by including gradient corrections can help to identify parameter regimes for which the TF solutions can be regarded as adequate.

\section{Quantum hydrodynamics with gradient corrections}\label{Quantumhydro}

One possibility to tackle interacting quantum many-body problems is based on the hydrodynamical form of the Schr\"odinger equation, first introduced by Madelung \cite{Madelung1927}. We shall use the inverse Madelung transformation to derive a nonlinear pseudo-Schr\"odinger equation for our many-fermion system, whose ground state candidates are then obtained by imaginary-time evolution (ITE). This approach quite naturally implies the variational equations (\ref{TFvWeq}) with and without gradient corrections of the kinetic energy functional, but circumvents their technically tedious direct solving. Most importantly, it offers the advantage of going beyond the TF equations by enforcing differentiability of the particle densities, while gradient corrections can be included simply by specifying $\xi$ in (\ref{DeltaT}). This enables us to study the impact and importance of the gradient corrections both qualitatively and quantitatively.

Introducing the pseudo-wave function
\begin{align}\label{TwoCompPsi}
\psi=\spinor{\psi_+}{\psi_-}{\phantom{\Delta\tau}}=\spinor{\sqrt{n_+}\,e^{\I\frac{m}{\hbar}\chi_+}}{\sqrt{n_-}\,e^{\I\frac{m}{\hbar}\chi_-}}{\phantom{\Delta\tau}},
\end{align}
we derive the Euler-Lagrange equations for the four fields $n_\pm$ and $\chi_\pm$ from the Hamiltonian
\begin{align}\label{HmanyFermions}
H=T_{\mathrm{tot}}+\int(\d\vec r)\Big\{V_+\,n_++V_-\,n_-+g\,n_+n_-\Big\}.
\end{align}
Here, ${n_++n_-=\psi^\dagger\psi}$ is the total one-particle density, and ${\nab\chi_\pm=\vec v_\pm}$ are the velocity fields of the collective motion of the flows $n_\pm$. We want to stress that $\psi$ is not a wave function, but merely a mean field that combines densities and velocity fields of the two fermion species. The spin-dependence of the external potentials $V_\pm$ enables us to break the spherical symmetry of isotropic potentials and study the stability of our numerical solutions in the limit ${V_+=V_-}$. The total kinetic energy ${T_{\mathrm{tot}}[n_\pm,\chi_\pm]=T+T_{\mathrm{c}}}$ in (\ref{HmanyFermions}) is composed of the intrinsic kinetic energy $T$ and the kinetic energy of the collective motion of the fermions, ${T_{\mathrm{c}}=\sum_{j=\pm}\int(\d\vec r)\frac{m}{2}n_j\,\vec v_j^2}$. With the functional derivatives ${\frac{\delta T_{\mathrm{c}}}{\delta n_\pm}=\frac{m}{2}\vec v_\pm^2}$ the four Euler-Lagrange equations read
\begin{align}\label{TwoCompMadelungHydro}
\begin{split}
\partial_t n_\pm&=-\nab(n_\pm\,\vec v_\pm),\\
m\partial_t\vec v_\pm&=-\nab\left(\frac{\delta T}{\delta n_\pm}+\frac{m}{2}\vec v_\pm^2+V_\pm+g\,n_\mp\right).
\end{split}
\end{align}

The hydrodynamical equations (\ref{TwoCompMadelungHydro}) are formally identical to Madelung's equations and generally describe fermions in motion. In the following we assume rotation-free velocity fields ${\vec v_\pm=\nab\chi_\pm}$. This condition is in particular fulfilled for stationary ground states, which obey ${\vec v_\pm=0}$. 

The inverse Madelung transformation amounts to explicitly recasting the time evolution $\I\hbar\partial_t\psi_\pm$ of the (pseudo-)wave function (\ref{TwoCompPsi}) in terms of the time-derivatives of the densities and phases from (\ref{TwoCompMadelungHydro}). For our many-body system we have to employ an approximation for the density-functional $T$, if we want to evaluate (\ref{TwoCompMadelungHydro}) numerically. Natural choices are the TF approximation (\ref{TTF}), with or without its corrections (\ref{DeltaT}). For 3D we get
\begin{align}\label{deltaT3D}
\frac{\delta T}{\delta n_\pm}\approx A\,n_\pm^{2/3}-\xi\frac{\hbar^2}{2m}\frac{\nab^2\sqrt{n_\pm}}{\sqrt{n_\pm}}
\end{align}
and obtain the pseudo-Schr\"odinger equation
\begin{align}\label{pseudopsi}
\I\hbar\partial_t\psi_\pm&\approx \Big[-\frac{\hbar^2}{2m}\nab^2+\frac{\hbar^2}{2m}(1-\xi)\frac{\nab^2|\psi_\pm|}{|\psi_\pm|}\nn\\
&\quad+A\,|\psi_\pm|^{4/3}+V_\pm+g |\psi_\mp|^2\Big]\psi_\pm.
\end{align}
The gradient corrections that reach beyond the usual TF-approximation are naturally included in the hydrodynamical approach via the dimensionless parameter $\xi$ and leave (\ref{pseudopsi}) structurally unchanged. In the following we argue that the hydrodynamical approach goes beyond the TF equations even when explicit gradient corrections are omitted.

The variational equations (\ref{TFvWeq}) for the stationary ground state are recovered from the hydrodynamical equations (\ref{TwoCompMadelungHydro}) in noting that ${\partial_t n_\pm=0}$ and ${\vec v_\pm=0}$: The two expressions ${\frac{\delta T}{\delta n_\pm}+V_\pm+g\,n_\mp}$ in (\ref{TwoCompMadelungHydro}) then equal some constants which we may identify with the chemical potentials $\mu_\pm$. That does not mean, however, that Madelung's differential equations (\ref{TwoCompMadelungHydro}) for ${\xi=0}$ are equivalent to the TF equations (\ref{TFeq}), where gradient corrections are omitted as well. The crucial difference between the TF-solutions and the densities from (\ref{TwoCompMadelungHydro}) for ${\xi=0}$ is differentiability --- which is lost in going from (\ref{TwoCompMadelungHydro}) to (\ref{TFeq}). Hence, even for ${\xi=0}$ we cannot expect the same solutions from (\ref{TwoCompMadelungHydro}) and (\ref{TFeq}). Retaining differentiability means, at the level of the energy functional and the variational equations, that gradient corrections have to be taken into account. In the remainder of this work we present stationary solutions of (\ref{pseudopsi}), \ie, candidates for the ground state of our many-body problem, obtained from the well-known imaginary-time evolution method.

\section{Results of imaginary-time evolution}\label{ImagTimeEvol}

The replacement ${t\to-\I\tau}$ in the time evolution of the linear Schr\"odinger equation enforces exponential decay of energy eigenstates with increasing real $\tau$. Then, the relative contribution from a non-degenerate ground state decays the slowest, such that an approximate ground state is obtained for large enough evolution time. Although the solutions of the nonlinear equation (\ref{pseudopsi}) do not represent actual wave functions, (\ref{pseudopsi}) is reminiscent of a (generalized) Gross-Pitaevskii equation, for which ITE has a very good track record, see \cite{Amara1993,Chiofalo2000,Baye2010,Antoine2014a} for some examples. In particular, the energy-diminishing property of ITE has been established rigorously, such that ITE converges to local energy minima \cite{Bao2004}. Also multi-component systems with coupled nonlinearities have been studied successfully \cite{Antoine2014a}.

The evolution of the pseudo-wave function towards stationarity can require a long propagation in imaginary time. Depending on the choice of the initial state, ITE can yield various stationary states that differ in energy only by small amounts, and deciding which is a better candidate for the ground state density is therefore not straightforward. One possibility for such ambiguities is metastability, which may then also be observed in the laboratory. To obtain unambiguous stationary densities we employ initial Gaussian states that are dressed with position-dependent noise on amplitude, mean, and width. Furthermore, we use slightly different external potentials for the two fermion components, ${V_\pm(\vec r)=V(\vec r)\mp\vec F\cdot\vec r}$, with the constant vector $\vec F$ not pointing along any symmetry axis of the numerical grid. From here onwards we use ${|\vec F|=10^{-6}}$, which should be well within experimental noise of a realistically applied potential $V(\vec r)$. The density profiles are essentially unchanged for different but small enough $|\vec F|$, indicating a stable numerical solution \wrt small changes of the potential.

Our main results are presented in the following. In Sec.~\ref{PhaseTransition} we encounter two types of separations and present the corresponding phase diagram of interaction strength versus particle number. We address the quantitative and qualitative importance of the gradient corrections in Sec.~\ref{GradCorr}. In Sec.~\ref{BarrierPotentials} we turn to more complex potentials by adding tunneling barriers in the center of the harmonic trap.

\subsection{Phase transitions}\label{PhaseTransition}

\begin{figure}[h!tbp]
\centering
\includegraphics[width=0.99\linewidth]{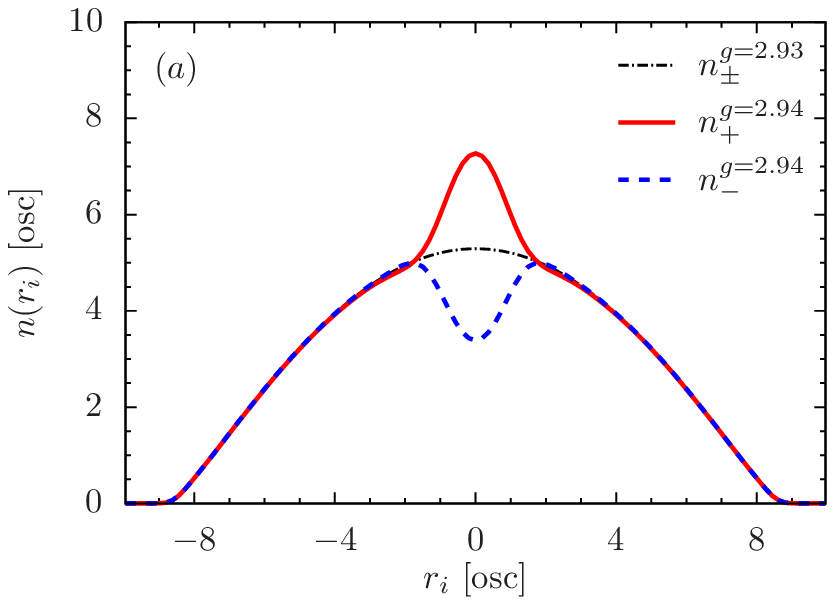}
\includegraphics[width=0.99\linewidth]{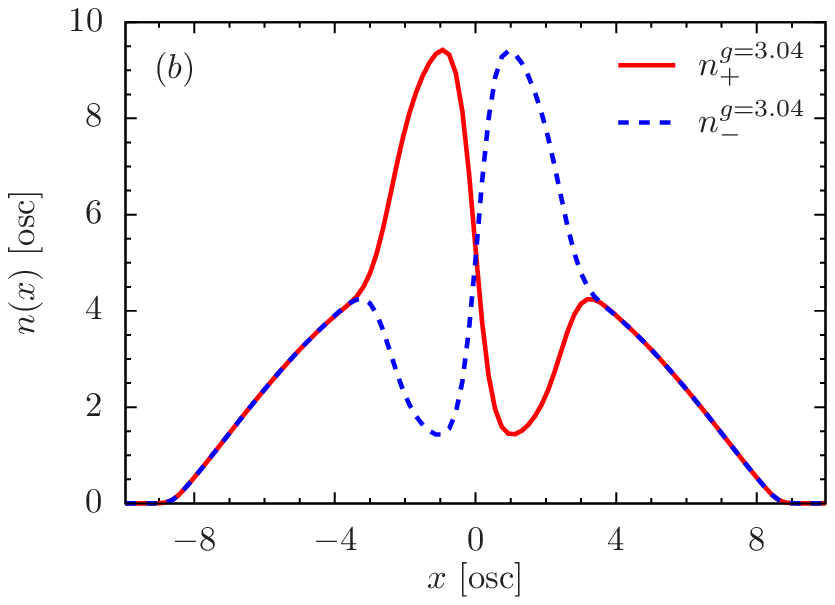}
\caption{(Color online) Two phase transitions for the particle densities $n_\pm$ with ${N_\pm=5000}$ in a harmonic trap are revealed as the repulsion strength $g$ grows. (a) The transition from non-separated densities towards isotropic separation occurs at $g_{\mathrm{is}}$, with ${2.93<g_{\mathrm{is}}<2.94}$. (b) The transition from isotropic towards anisotropic separation occurs at $g_{\mathrm{as}}$, with ${3.03<g_{\mathrm{as}}<3.04}$.}
\label{ExampleDensities20150619}
\end{figure}

\begin{figure}[h!tbp]
\centering
\includegraphics[width=0.99\linewidth]{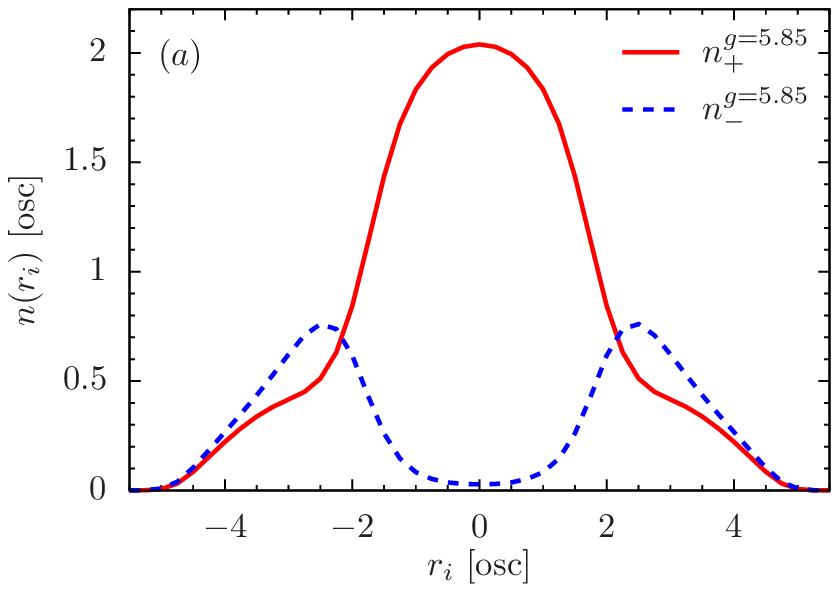}
\includegraphics[width=0.99\linewidth]{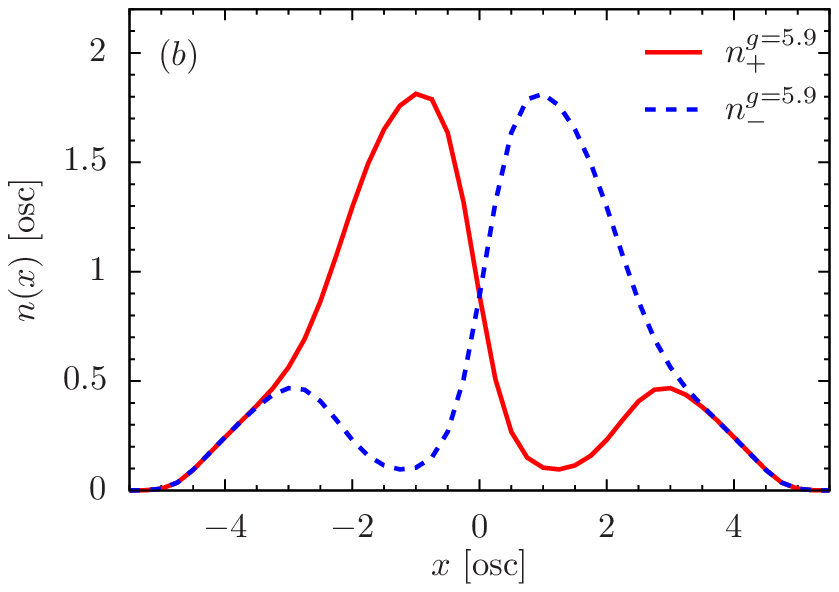}
\includegraphics[width=0.99\linewidth]{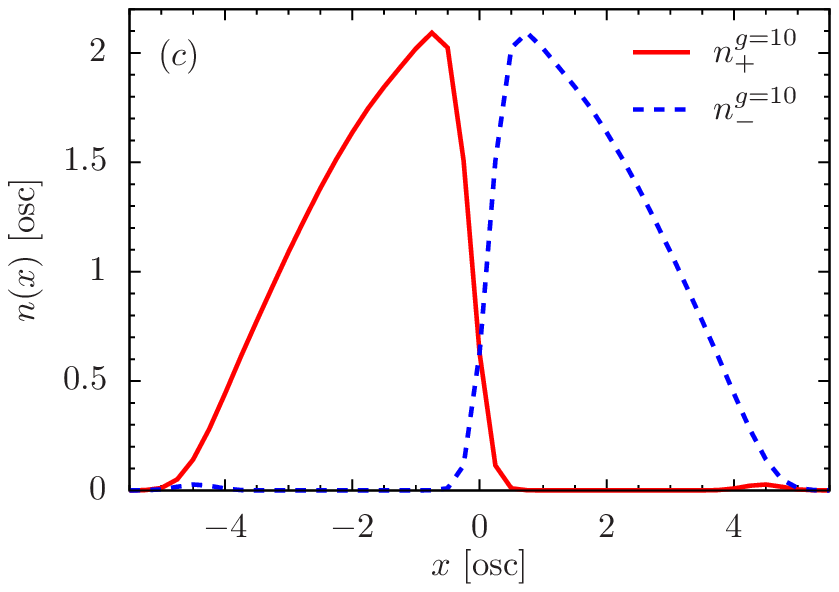}
\caption{(Color online) The densities for ${N_\pm=160}$ in a harmonic trap for various $g$, illustrating the two phase transitions (cf.~Fig.~\ref{ExampleDensities20150619}). (a)/(b) A transition from isotropic to anisotropic separation into two 'semi-spheres' is encountered between ${g=5.85}$ and ${g=5.9}$, indicating spontaneous symmetry breaking as soon as the depletion of one component in the center is total. (c) With $g$ well beyond the second phase transition, we gain a quasi complete separation of the two Fermi gas clouds.}
\label{2ndPhaseTrans160}
\end{figure}

In this section we present our results for phase transitions which occur as the repulsion between the two Fermi components in a harmonic oscillator potential ${V(\vec r)=r^2/2}$ is increased. We find a sharp phase transition from symmetric density profiles (${n_+(\vec r)=n_-(\vec r)}$ for all $\vec r$) to a partial separation of the two fermion species once a critical interaction strength is exceeded. Increasing the repulsion further, we find a second phase transition from the spherically symmetric separation towards an anisotropic split, such that the two species almost completely separate for large $g$. A phase transition from symmetric to isotropically separated densities (${n_+(\vec r)\not=n_-(\vec r)}$ for some $\vec r$) is observed at a critical interaction strength ${g_{\mathrm{is}}=g_{\mathrm{is}}(N_+=N_-)}$, see Fig.~\ref{ExampleDensities20150619}(a). One component is partially depleted in the center of the harmonic trap and pushed outwards, while still retaining spherical symmetry. With increasing $g$ the isotropic separation becomes more pronounced and a transition towards an anisotropic splitting of the two fermion species is observed for ${g>g_{\mathrm{as}}=g_{\mathrm{as}}(N_+=N_-)}$, as illustrated in Figs.~\ref{ExampleDensities20150619}(b) and \ref{2ndPhaseTrans160}. For even larger repulsion, the densities tend to repel each other stronger, until no appreciable overlap is found.


\begin{figure}[h!tbp]
\centering
\begin{minipage}{\linewidth}
\begin{minipage}{0.44\linewidth}
\includegraphics[height=15em]{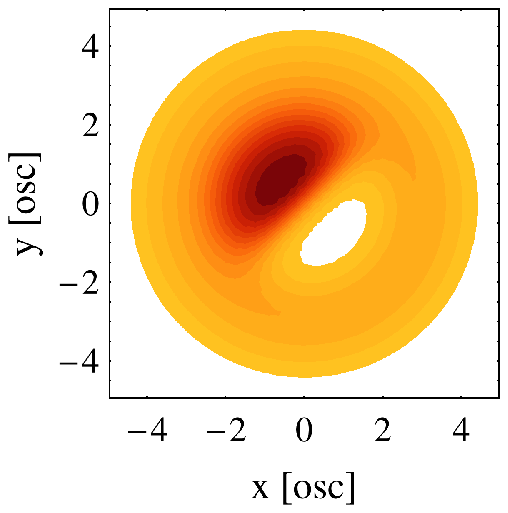}
\end{minipage}
\hfill
\begin{minipage}{0.44\linewidth}
\includegraphics[height=15em]{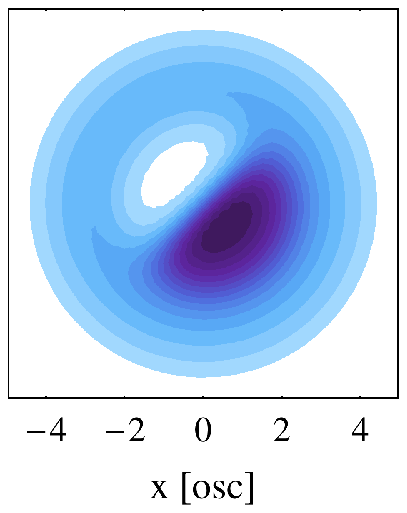}
\end{minipage}
\end{minipage}
\caption{(Color online) Contour plots of the normalized densities ${\hat{n}_\pm(\vec r)=\hat{n}_\pm(\vec r)/\mathrm{max}_{\vec r}n_\pm(\vec r)}$ (red/blue) in the ${(z=0)}$-plane for ${N_\pm=160}$ and ${g=5.9}$ with color gradient from white (${\hat{n}_\pm(\vec r)<0.01)}$) to darkest color (${\hat{n}_\pm(\vec r)>0.93)}$).  The density profiles along the $x$-axis are displayed in Fig.~\ref{2ndPhaseTrans160}(b).}
\label{20150617TFWcontourPlotxy}
\end{figure}

\begin{figure}[h!tbp]
\centering
\includegraphics[width=0.99\linewidth]{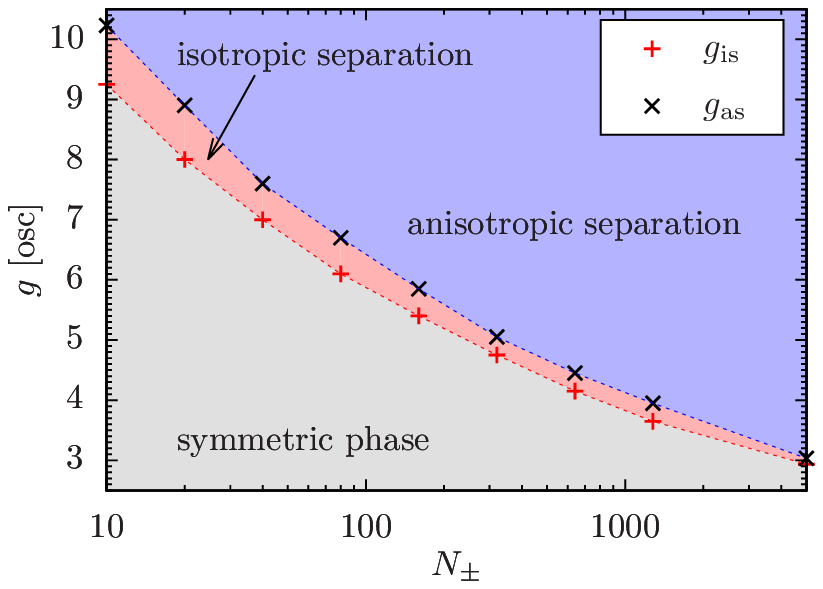}
\caption{(Color online) Phase diagram of critical repulsion strengths $g_{\mathrm{is}}$ and $g_{\mathrm{as}}$ as functions of the particle numbers ${N_+=N_-}$ for the harmonic oscillator potential ${V=r^2/2}$.}
\label{gcrit}
\end{figure}

Figure~\ref{20150617TFWcontourPlotxy} shows representative contour plots of $n_\pm$ in the ${(z=0)}$-plane for an isotropic harmonic trap. The numerics yields a state with spontaneously broken symmetry, a splitting into two semi-spheres. Of course, the direction of the splitting for isotropic harmonic confinement may occur in any spatial direction. In contrast, anisotropies enforce a specific axis of separation. For example, the interface between $n_+$ and $n_-$ lies in the ${(z=0)}$-plane if an elongated harmonic trap with smallest frequency in $z$-direction is employed. In general, our numerical data suggest that a minimal interface between the two Fermi components is energetically preferred.

The dependences of the critical interaction strengths $g_{\mathrm{is}}$ and $g_{\mathrm{as}}$ on the particle numbers $N_\pm$ are illustrated with the phase diagram in Fig.~\ref{gcrit}. For all particle numbers considered we find a transition from the symmetric phase to isotropic separation and from isotropic to anisotropic separation. While both $g_{\mathrm{is}}$ and $g_{\mathrm{as}}$ decrease for increasing $N_\pm$, the range of $g$ that allows for an isotropic separation shrinks \footnote{Although the ITE method is generally not restricted by the number of particles $N_\pm$, the available computing resources limit an efficient investigation for such large $N_\pm$. Our numerical data for ${N_\pm=500000}$ indicate that the trend of the critical interaction strengths as functions of $N_\pm$ displayed in Fig.~\ref{gcrit} continues beyond the shown particle numbers.}. We shall observe similar phase transitions for various trap geometries. In the following section we investigate the qualitative and quantitative impact of the gradient corrections on the density profiles.

\subsection{Importance of gradient corrections}\label{GradCorr}

One main result of this work is the necessity to go beyond the TF-approximation to obtain viable stationary states even qualitatively. For ${\xi=0}$ the magnitude of the gradient corrections (\ref{DeltaT}) vanishes. However, differentiability is still retained within the formalism of ITE --- as opposed to the TF equations, which correspond to ${\xi=0}$ as well, but do not lead to differentiable densities. For interaction strengths near the two phase transitions discussed in Sec.~\ref{PhaseTransition} we calculate the densities for ${\xi=0}$ and compare with the case of ${\xi=1/9}$. We find that gradient corrections can actually be relatively large and are therefore relevant for a quantitative analysis of the particle densities. Although the TF density can be reasonably accurate for selected parameters, cf.~Fig.~\ref{ExampleDensities20150601b} in the appendix, we argue that a systematic inclusion of differentiability is generally required.

\begin{figure}[ht!]
\centering
\includegraphics[width=0.99\linewidth]{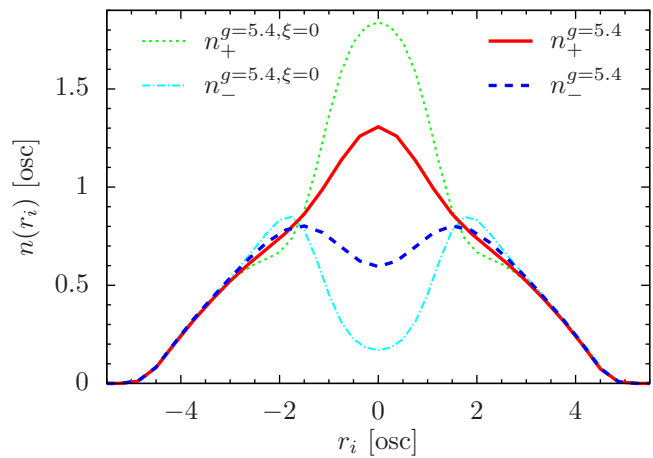}
\caption{(Color online) Densities for ${N_\pm=160}$ in the harmonic oscillator potential ${V=r^2/2}$ for ${g=5.4}$ near the phase transition from symmetric profiles to isotropic separation.}
\label{ExampleDensities20150601}
\end{figure}

\begin{figure}[ht!]
\centering
\includegraphics[width=0.99\linewidth]{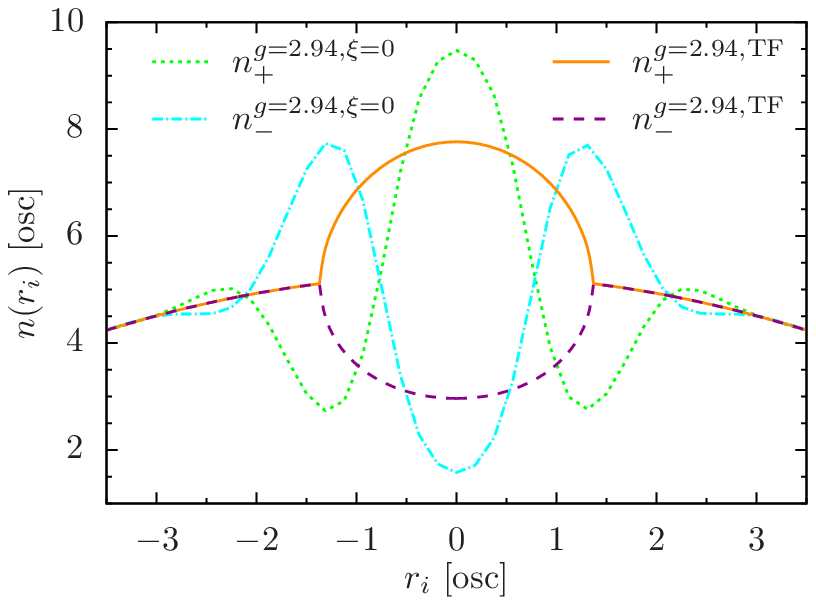}
\includegraphics[width=0.99\linewidth]{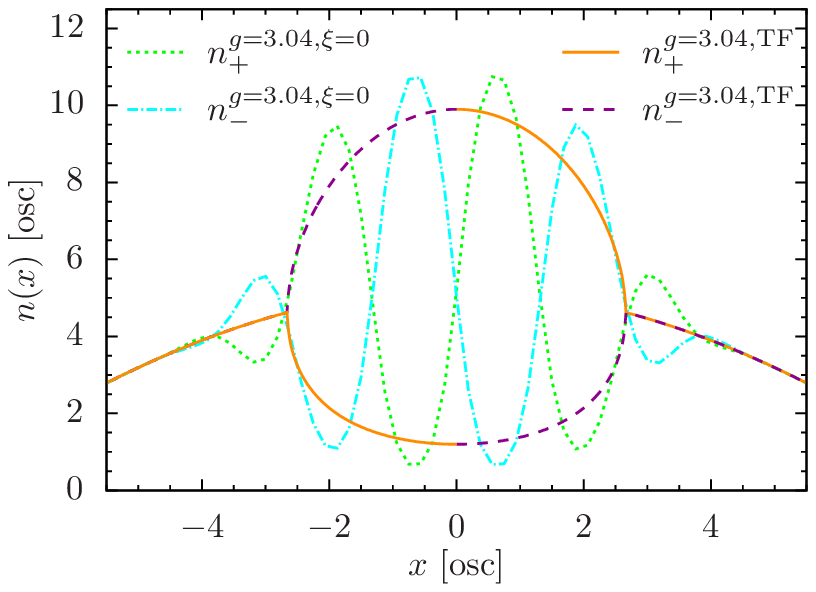}
\caption{(Color online) Densities for ${N_\pm=5000}$ fermions in the harmonic oscillator potential ${V=r^2/2}$. Both the TF solutions (with ${\mu_+=\mu_-}$) and the densities obtained from ITE show an isotropic separation in the trap center for ${g=2.94}$ (upper figure), whereas an anisotropic separation for ${g=3.04}$ is only obtained by ITE, unless we disregard continuity for the TF solutions (lower figure).}
\label{ExampleDensities20150601bg2dot94}
\end{figure}
\begin{figure}[ht!]
\centering
\includegraphics[width=0.99\linewidth]{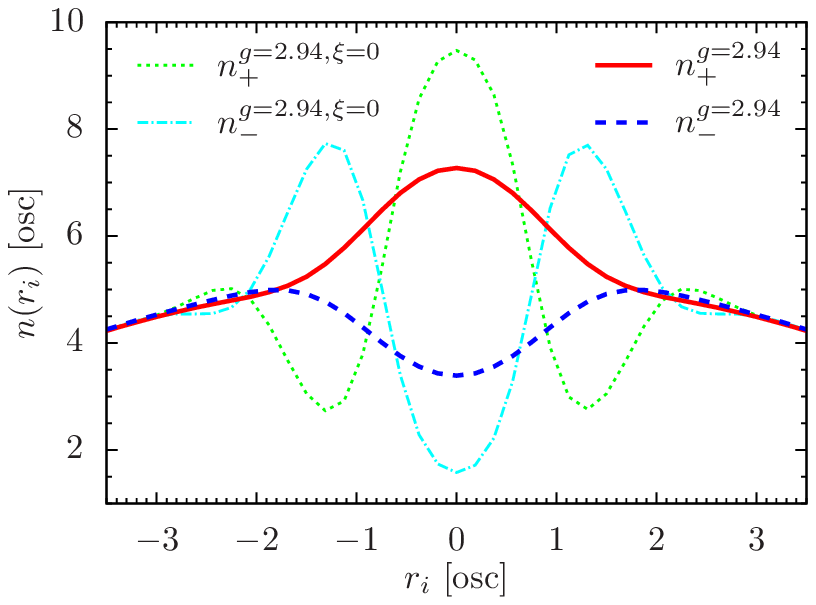}
\includegraphics[width=0.99\linewidth]{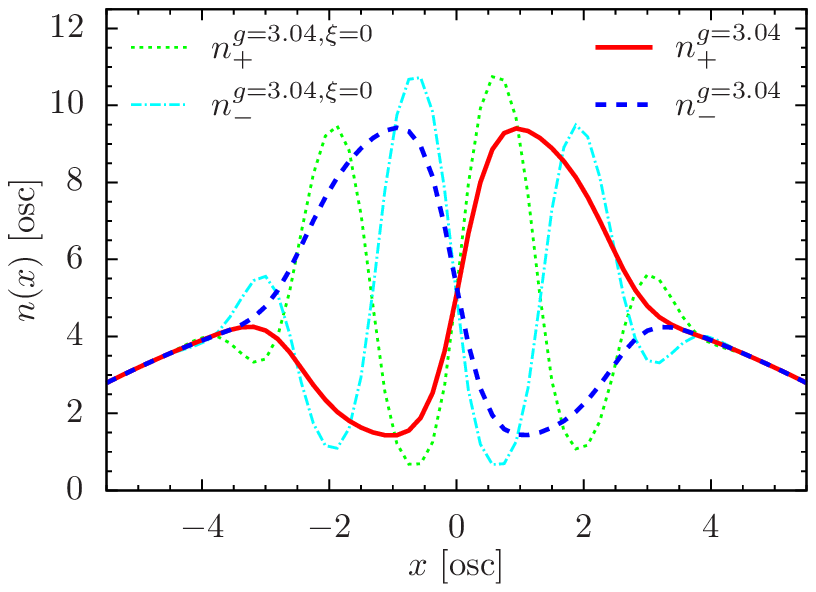}
\caption{(Color online) Densities for ${N_\pm=5000}$ from ITE with (${\xi=1/9}$) and without (${\xi=0}$) explicit gradient corrections. We observe isotropic separation for ${g=2.94}$ (upper figure) and anisotropic separation for ${g=3.04}$ (lower figure).}
\label{ExampleDensities20150601g2dot94}
\end{figure}

At the example of ${N_\pm=160}$ and for $g$ close to $g_{\mathrm{is}}$, we demonstrate in Fig.~\ref{ExampleDensities20150601} that gradient corrections have to be considered quantitatively important for the computation of density profiles. For ${N_\pm=5000}$ the effects of gradient corrections are even more pronounced. Figures~\ref{ExampleDensities20150601bg2dot94} and \ref{ExampleDensities20150601g2dot94} show density profiles for $g$ close to $g_{\mathrm{is}}$ and $g_{\mathrm{as}}$, respectively. The densities shown in Fig.~\ref{ExampleDensities20150601bg2dot94} are obtained by employing the analytical TF solutions discussed in the appendix and by using ITE with ${\xi=0}$, respectively. Although the global features are similar, the TF densities differ significantly in spatial regions where separations of $n_+$ and $n_-$ appear, as opposed to the results of ITE(${\xi=0}$) shown in Fig.~\ref{ExampleDensities20150601b} in the appendix. Figure~\ref{ExampleDensities20150601g2dot94} demonstrates that explicit gradient corrections have a qualitative impact on the density profiles even within the ITE method since the result of ITE with ${\xi=0}$ exhibits significant qualitative differences to ITE(${\xi=1/9}$). This suggests that not only differentiability, that is, gradients of the densities, but also the magnitudes of the gradient corrections are relevant for the qualitative features of the density profiles. In the next section we investigate barrier potentials that may be regarded as prototypes of external potentials that are used in recent experiments like \cite{Roati2011}.

\subsection{Density profiles for barrier potentials}\label{BarrierPotentials}

For an isotropic Gaussian potential at the center of the harmonic trap, \ie, ${V(r)=0.5\,r^2+\nu\times\exp\big(-20\,r^2\big)}$, where ${\nu=300}$, we again find a phase transition from symmetric to spherically separated densities. The densities arrange spherically around the central Gaussian, which is about a factor of $6$ higher than the chemical potentials. For $g$ between 3.05 and 3.1 we encounter the phase transition towards anisotropic separation, similar to the separations illustrated in Figs.~\ref{2ndPhaseTrans160}(b) and \ref{20150617TFWcontourPlotxy}.

\begin{figure}[h!tbp]
\centering
\includegraphics[width=0.99\linewidth]{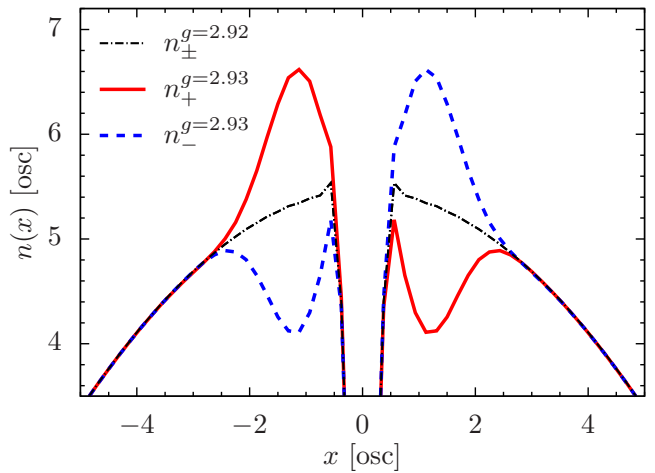}
\caption{(Color online) Employing a barrier at ${x=0}$ that marginally exceeds the chemical potentials, we observe a direct transition from a symmetric state to an anisotropic separation --- very similar to the case of a higher barrier --- between ${g_1=2.92}$ and ${g_1=2.93}$.}
\label{ExampleDensities20150813}
\end{figure}

We test the ability of the ITE method to deal with tunneling processes that are in principle included in our description through gradient corrections by adding a barrier in $x$-direction instead of an isotropic Gaussian potential, \ie, ${V(r)=0.5\,r^2+\nu\times\exp\big(-20\,x^2\big)}$. In this situation of two classically separated wells we find that the symmetry-breaking barrier at ${x=0}$ drives the fermion clouds from a symmetric state ${\big(n_+(\vec r)=n_-(\vec r)\big)}$ directly into an anisotropic separation ${\big(n_+(x)\not=n_-(x)}$, ${n_+(y)=n_-(y)}$, ${n_+(z)=n_-(z)\big)}$ as $g$ increases (\eg, from ${g=2.91}$ to ${g=2.92}$ for ${N_\pm=5000}$ and ${\nu=300}$). Virtually the same profiles are obtained for ${N_\pm=5000\pm0.01}$ and ${N_\pm=5000\pm1}$. Of course, in the two latter cases the densities of the differently occupied components do not coincide entirely even for ${g=2.91}$. The excess of $N_+$ shows up symmetrically on both half-axes in $x$-direction.

Our results for a lowered barrier are displayed in Fig.~\ref{ExampleDensities20150813}, where we choose ${\nu=60}$ and find a transition from symmetric profiles to anisotropic separation in $x$-direction if $g$ is increased from ${g_1=2.92}$ to ${g_2=2.93}$. As in the case of ${\nu=300}$ the profiles in $y$- and $z$-direction are virtually the same for $g_1$ and $g_2$.

\clearpage

\section{Conclusions and perspectives}

In this article we discussed the importance of gradient corrections for an adequate description of two-component Fermi gases with repulsive contact interaction and focused on possible separations of the two fermion species. We demonstrated that gradient corrections beyond the Thomas-Fermi approximation are crucial even for the qualitative features of the ground-state particle densities. We obtained candidates for ground state densities in three dimensions via imaginary-time evolution of a pseudo-Schr\"odinger equation that we derived with the aid of an inverse Madelung transformation. Since the gradient corrections enter this hydrodynamical formulation naturally in terms of a single parameter, we were able to study the impact of the gradient corrections both qualitatively and quantitatively within the same approach.

Our numerical results for particle numbers up to 10000 revealed two phase transitions. While the densities of the two fermion species are the same and constitute a symmetric phase for small repulsion $g$, they start to separate once a critical interaction strength is exceeded. For isotropic harmonic confinement one of the Fermi components is isotropically repelled from the trap center. For even larger $g$ we found a second phase transition towards an anisotropic separation into two semi-spheres, such that no appreciable overlap of the two components remains for very large $g$. We established our method in view of experiments on ultracold Fermi gases, for which more complex trapping potentials have to be considered. For example, adding a symmetry-breaking tunneling barrier in the trap center, we found the anisotropic phase to emerge directly from the symmetric phase as $g$ is increased beyond a critical value. The results shown in this work suggest that imaginary-time evolution of the pseudo-Schr\"odinger equation (\ref{pseudopsi}) is a viable tool to obtain gradient-corrected candidates of ground-state densities for contact-interacting two-component Fermi gases in any realistic (bounded) external potential.

The here developed formalism allows us to use the ground-state pseudo-wave functions as approximate ground states and to study dynamics simply by switching from imaginary- to real-time evolution of the pseudo-Schr\"odinger equation. Further directions of investigation are certainly experimentally valuable extensions like the inclusion of dipole-dipole interactions. It would also be interesting to use the formalism developed in this work for low-dimensional systems that are numerically more tractable than the three-dimensional case, as soon as corresponding gradient corrections of the kinetic energy functional are available.

\begin{acknowledgments}
We want to express our gratitude to B.-G.~Englert, M.~Gajda, B.~Gr\'emaud, T.~Karpiuk, K.~Paw{\l}owski and T.~Sowi\'nski for valuable discussions. This work was supported by the (Polish) National Science Center Grant No.~DEC-2012/04/A/ST2/0009.
\end{acknowledgments}

\appendix*

\section{Thomas-Fermi densities}

In this appendix we discuss some properties of the TF equations (\ref{TFeq}), whose solutions represent local extrema of $E_0[n_+,n_-,\mu_+,\mu_-]$, see (\ref{Etot}). In the noninteracting case (${g=0}$) the two TF equations decouple, and their physically valid solutions are the nonnegative particle densities ${n_\pm=\big[(\mu_\pm-V)/A\big]^{D/2}\,\Theta(\mu_\pm-V)}$. The solutions of (\ref{TFeq}) are restricted to the classically allowed regions given by ${\mu_\pm-V> 0}$. The Heaviside step function  ${\Theta(\cdot)}$ enforces $n_i=0$ wherever the solution of the TF equation ${\frac{\delta E_0}{\delta n_i(\vec r)}=0}$ does not yield a nonnegative value $n_i$. We may call this density for all space the TF-density $n^{\mathrm{TF}}(\vec r)$.

In contrast to the TF energy functional $E_0$ whose support in the function space may include vanishing densities, the support of the TF equations is restricted to ${n_\pm>0}$ (for any $g$) because variations ${n_i(\vec r)-|\delta n_i(\vec r)|}$ (being negative if ${n_i(\vec r)=0}$) of $E_0$ are not permissible, such that the TF equation for the component $n_i$ is not valid at $\vec r$ in the first place \footnote{Following the same argumentation, we exclude complex variations since all test densities have to integrate to a real particle number.}. Whether or not either of the TF equations is valid at a given position has to be decided for each $\vec r$ independently. If both equations (\ref{TFeq}) are valid at some $\vec r$, they may be solved self-consistently for $n_\pm(\vec r)$. If, however, one component $n_i(\vec r)$ becomes zero at some $\vec r$ (i.e.~at the quantum classical boundary for the component $i$) or complex, a variation of $E_0$ \wrt $n_i(\vec r)$ would require to leave the support of $E_0$. In this case, only ${\frac{\delta E_0}{\delta n_j(\vec r)}=0}$ ${(j\not=i)}$ may be used to determine $n_j(\vec r)$, which then corresponds to the value $n_j(\vec r)$ of the noninteracting solution since ${n_i(\vec r)=0}$.

\begin{figure}[h!tbp]
\centering
\includegraphics[width=0.99\linewidth]{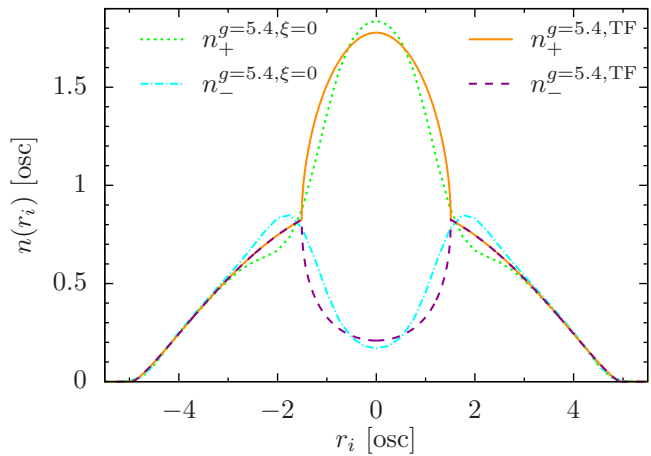}
\caption{(Color online) Isotropic densities for ${N_++N_-=320}$ in the harmonic oscillator potential ${V(\vec r)=\vec r^2/2}$. The ITE (without explicit gradient corrections) is employed with ${N_\pm=160}$ for ${g=5.4}$. Disregarding the deviating particle numbers ${N_\pm^{\mathrm{TF}}=160\pm 7}$, we find that the TF-approximation works reasonably well at the qualitative level.}
\label{ExampleDensities20150601b}
\end{figure}

In general, (\ref{TFeq}) yield a myriad of possible density profiles since the TF equations at different positions $\vec r$ are decoupled and can be combined in many different ways. Since we cannot get rid of this ambiguity, we introduced gradient corrections to find ground state candidates in Sec.~\ref{Quantumhydro}. However, it turns out in retrospect that the TF equations, when restricted to continuous densities, actually predict reasonable density profiles for selected parameters similar to the more sophisticated approach of ITE, see Fig.~\ref{ExampleDensities20150601b}. For smooth isotropic potentials ${V=V(r)}$ that are strictly monotonously increasing in the radial variable ${r=|\vec r|}$, $\mu_+=\mu_-$, and $g>0$ we obtain analytical solutions in all dimensions by investigating the sum and difference of the TF equations. However, we show that it is in general impossible to obtain the global minimum of ${E_0[n_+,n_-,\mu_+,\mu_-]}$ from these restricted TF equations.

Unless $g$ is unrealistically fine-tuned to $A$, we obtain the 2D solutions of (\ref{TFeq})
\begin{align}\label{n2D}
n_+=n_-=\frac{\mu-V}{A+g}\Theta(\mu-V),
\end{align}
which enforces ${N_+=N_-}$ and provides no separations.

With ${\rho_\pm^3=n_\pm}$ we find for 3D
\begin{align}
\rho_+^2+\rho_-^2=\big[2(\mu-V)-g(\rho_+^3+\rho_-^3)\big]/A\label{TFeq3Da}
\end{align}
and
\begin{align}
(\rho_+-\rho_-)\big[\rho_++\rho_--g(\rho_+^2+\rho_-^2+\rho_+\rho_-)/A\big]=0,\label{TFeq3Db}
\end{align}
where (\ref{TFeq3Db}) implies either a symmetric (${\rho_\pm=\rho}$) or a separated (${\rho_+\not=\rho_-}$) solution. The symmetric solution is disjunct from the separated solution in the sense that they cannot hold simultaneously on any finite spatial interval since this would require constant $\rho^3$, \ie, constant $V$. Since $g$, $A$, and ${\mu-V}$ are positive, there is always exactly one physically valid symmetric solution, obtained from (\ref{TFeq3Da}), namely the (only) non-negative real root of ${\rho^3+A/g\,\rho^2-(\mu-V)/g=0}$. With (\ref{TFeq3Da}) and (\ref{TFeq3Db}), we find the separated solutions from ${y=x(x-a)}$ and ${P(x)=0}$, where ${x=\rho_++\rho_-}$, ${y=\rho_+\rho_-}$, ${a=A/g}$, and ${P(x)=x^3-a\,x^2-a^2x+a^2\,g(\mu-V)/A^2}$. From ${y>0}$ we immediately see that ${x\notin (0,a)}$, \ie, the symmetric solution is always part of the entire density profile near $r_0$, where ${x=0}$. There is always exactly one positive root $x_P$ of $P(x)$ with ${x_P\ge a}$. Thus,
\begin{align}\label{npmTF3D}
n_\pm=\left[\frac{x_P}{2}\left(1\pm\sqrt{\frac{4a}{x_P}-3}\right)\right]^3\Theta(\mu-V),
\end{align}

The generic structure of the TF solutions in 1D is similar to the 3D case, cf.~Fig.~\ref{TwoFermiComponentsPlotsn3D},
\begin{align}
n&=\frac{-g}{2A}\left(1-\sqrt{\frac{4A}{g^2}(\mu-V)+1}\right)\Theta(\mu-V),\label{1Dsymm}\\
n_\pm&=\frac{g}{2A}\left(1\pm\sqrt{\frac{4A}{g^2}(\mu-V)-3}\right)\Theta(\mu-V) .\label{1Dsplit}
\end{align}

Separated solutions are obtained in 1D if and only if ${\sqrt{A\mu}<g<\sqrt{4A\mu/3}}$. The according interval in 3D, ${\sqrt{20A^3/(27\mu)}<g<\sqrt{A^3/\mu}}$, shrinks with increasing $\mu$, that is, with increasing $N$. This observation is reminiscent of the phase of isotropic separation depicted in Fig.~\ref{gcrit}. Furthermore, there is only a single point where the separated solution merges with the symmetric solution, determined by ${V(r_1)=\mu-3g^2/(4A)}$ in 1D and ${V(r_1)=\mu-20A^3/(27g^2)}$ in 3D. We therefore conclude ${N_+=N_-\Leftrightarrow n_+=n_-}$ from the TF equations in any dimension for a monotonously increasing isotropic potential if ${\mu_+=\mu_-}$, provided that we demand continuous densities. Moreover, there are restrictions on the ratio ${N_+/N_-}$ and the range of interaction strengths $g$ for which separated solutions exist.

\begin{figure}
\centering
\vspace{1.5em}
\includegraphics[width=0.99\linewidth]{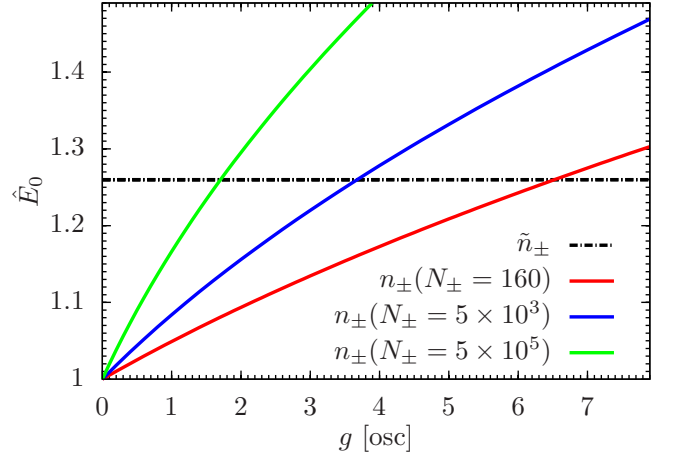}
\caption{(Color online) The normalized total TF-energies ${\hat E_0[n_\pm]=E_0[n_\pm;g]/E_0[n_\pm;g=0]}$ of symmetric solutions ${n_+=n_-}$ in 3D as a function of interaction strength $g$ for several $N_\pm$ (solid colored lines). The intersection with the constant value ${\hat E_0[\tilde{n}_\pm]}$ (dash-dotted line), obtained from the completely separated densities ${\tilde{n}_\pm=n_\pm(2N_\pm)\Theta(\pm x)}$ serves as an estimate of the critical strengths ${g^{\mathrm{TF}}(N_\pm)}$ above which symmetric solutions of the TF equations cannot represent the ground state.}
\label{TFEnergyListPlot}
\end{figure}

These findings are certainly counter-intuitive, and results from these continuous densities should be considered with caution. Indeed, with these restrictions the TF equations generally cannot yield the ground state densities. For large enough repulsion $g$, completely separated densities of the two fermion species in general give lower energies than the symmetric solutions from (\ref{TFeq}). This can be proven straightforwardly for the 2D harmonic oscillator. An analogous numerical calculation for 3D is illustrated in Fig.~\ref{TFEnergyListPlot}. We find ${g^{\mathrm{TF}}\approx6.5\,(3.6)}$ for ${N_\pm=160\,(5\times10^3)}$. In Sec.~\ref{PhaseTransition} we observed the onset of separations of the gradient-corrected fermion clouds at similar values of $g$, cf.~Figs.~\ref{ExampleDensities20150619} and \ref{2ndPhaseTrans160}.

\newpage

\bibliography{myPostDocbib20151113}

\end{document}